\documentclass{article}
\usepackage[T1]{fontenc}
\usepackage{wrapfig}
\usepackage[portuges,english]{babel}
\usepackage{amssymb,latexsym,amsmath,color,mathrsfs,ifsym,graphics,stmaryrd} %stmaryrd for \llbracket and \rrbracket
\usepackage[colorlinks,linkcolor=blue,urlcolor=blue,citecolor=black,
plainpages=false,pdfpagelabels,breaklinks]{hyperref}
\usepackage{graphicx}

\title{{\sc Mythical Thought in Bohr's\\Anti-Realist Realism\\(Or: Lessons on How to Capture\\ and Defeat Smoky Dragons)}}

\author{{\sc Christian de Ronde}\thanks{This paper is dedicated to the memory of Diego Armando Maradona who taught me, when I was a very young child, that the impossible can be achieved.}
}\date{}

\usepackage[margin=2.25 cm]{geometry}

\begin{document}
\maketitle

\begin{center}
\begin{small}
Philosophy Institute Dr. A. Korn, Buenos Aires University - CONICET\\ 
Engineering Institute - National University Arturo Jauretche, Argentina.\\
Federal University of Santa Catarina, Brazil.\\ 
Center Leo Apostel for Interdisciplinary Studies, Brussels Free University, Belgium.
\end{small}
\end{center}

\begin{abstract}
\noindent In this work we argue that the power and effectiveness of the Bohrian approach to quantum mechanics is essentially grounded on an inconsistent form of anti-realist realism which supports not only the uncritical tolerance ---in physics--- towards the ``standard'' account of the theory of quanta but also ---in philosophy--- to the mad reproduction of mythical (inconsistent and vague) narratives  ---known as ``interpretations''. We will discuss the existence of ---what John Archibald Wheeler named--- ``smoky dragons'' not only within the standard formulation of the theory but also within the many interpretations that have been ---later on--- introduced by philosophers and philosophically inclined physicists. After analyzing the role of smoky dragons within both contemporary physics and philosophy of physics we will propose a general procedure grounded on a series of {\it necessary theoretical conditions} for producing meaningful physical concepts that ---hopefully--- could be used as tools and weapons to capture and defeat these beautiful and powerful mythical creatures. 
\end{abstract}

%--------------------------------------------------------------
\renewenvironment{enumerate}{\begin{list}{}{\rm \labelwidth 0mm
\leftmargin 0mm}} {\end{list}}

\newcommand{\ita}{\textit}
\newcommand{\mcal}{\mathcal}
\newcommand{\mfrak}{\mathfrak}
\newcommand{\mbb}{\mathbb}
\newcommand{\mrm}{\mathrm}
\newcommand{\msf}{\mathsf}
\newcommand{\mscr}{\mathscr}
\newcommand{\lra}{\leftrightarrow}
\renewenvironment{enumerate}{\begin{list}{}{\rm \labelwidth 0mm
\leftmargin 5mm}} {\end{list}}

\newtheorem{dfn}{\sc{Definition}}[section]
\newtheorem{thm}{\sc{Theorem}}[section]
\newtheorem{lem}{\sc{Lemma}}[section]
\newtheorem{cor}[thm]{\sc{Corollary}}
\newcommand{\Proof}{\textit{Proof:} \,}
\newcommand{\cqd}{{\rule{.70ex}{2ex}} \medskip}

\bigskip 

\bigskip 

\bigskip 

\bigskip 

\bigskip 

\begin{flushright}
{\it Ts'ui Pe must have said once: I am withdrawing to write a book.\\
And another time: I am withdrawing to construct a labyrinth.\\ 
Every one imagined two works; to no one did it occur\\
that the book and the maze were one and the same thing.}\\
\smallskip
\smallskip
Jorge Luis Borges. 
\end{flushright}

\section{Realism and Anti-Realism in Physics}

Together with philosophy, physics was born born during the 6th century B.C. in the Greek city of Miletos, on the coast of Asia Minor, where the Ionians had established rich and prosperous colonies. Three men ---Thales, Anaximander, and Anaximenes--- appeared in quick succession claiming the existence of what they named {\it physis} ---translated later on as `reality'. For the first time in the history of western thought physicists replaced mythical stories and narratives by rational explanation. As remarked by Jean-Pierre Vernant:
\begin{quotation}
\noindent {\small ``Myths were accounts, not solutions to problems. They told of the sequence of actions by which the king or the god imposed order, as these actions were mimed out in ritual. The problem found its solution without ever having been posed. However, in Greece, where the new political forms had triumphed with the development of the city, only a few traces of the ancient rituals remained, and even their meaning had been lost. The memory of the king as creator of order and maker of time had disappeared. The connection is no longer apparent between the mythical exploit of the sovereign, symbolized by his victory over the monster, and the organization of cosmic phenomena. When the natural order and atmospheric phenomena (rains, winds, storms and thunderbolts) become independent from the functions of the king, they cease to be intelligible in the language of myth in which they had been described hitherto. They are henceforth seen as questions open for discussion. These questions (the genesis of the cosmic order and the explanation for {\it meteora}), in their new form as problems, constitute the subject matter for the earliest philosophical thought. Thus the philosopher takes over from the old king-magician, the master of time. He constructs a theory to explain the very phenomena that in times past the king had brought about.'' \cite[p. 402]{Vernant06}} 
\end{quotation}
The impact of this shift from myth to {\it theory} would reinforce the transformation of the structure of the Greek society and as a consequence, preachers, mediums and kings would be forced to share their power with the first physicists and philosophers. Unlike in the previous mythical society, in this new democratic system known as science, all citizens would be entitled to openly debate and gain theoretical understanding about reality:
\begin{quotation} 
\noindent {\small ``[Before the Milesians,] [e]ducation was based not on reading written texts but on listening to poetic songs transmitted from generation to generation. [...] These songs contained everything a Greek had to know about man and his past ---the exploits of heroes long past; about the gods, their families, and their genealogies, about the world, its form, and its origin. In this respect, the work of the Milesians is indeed a radical innovation. Neither singers nor poets nor storytellers, they express themselves in prose, in written texts whose aim is not to unravel a narrative thread in the long line of a tradition but to present an explanatory theory concerning certain natural phenomena and the organization of the cosmos. In this shift from the oral to the written, from the poetic song to prose, from narration to explanation, the change of register corresponds to an entirely new type of investigation ---new both in terms of its object (nature, {\it physis}) and in terms of the entirely positive form of thought manifested in it.'' \cite[p. 402]{Vernant06}} 
\end{quotation}

Three fundamental presuppositions would guide this new form of thought and {\it praxis}. First, that {\it physis} is not {chaotic}, that it possesses an internal order, what the Greeks called ---after Heraclitus--- a {\it logos}. Second, that {\it physis} is one. Like Heraclitus would express: ``Listening not to me but to the {\it logos} it is wise to agree that all things are one'' [f. 50 DK]. Third, that even though ``{\it physis} loves to hide'' [f. 123 DK] the logos of physis could be actually known through the development of {\it theories}, namely, through the creation of unified, consistent and coherent schemes of thought that could explain phenomena. Already the first philosophers showed a clear recognition of the difficult problems involved within this scheme. To relate the {\it logos} of men understood as discourse to the {\it logos} of {\it physis} was obviously a difficult task, but through hard work and sensibility the latter could be revealed in the former.\footnote{In this respect, it is important to recognize that physicists never claimed that {\it theories} ``mirrored'' reality. In fact, the idea that realism implies a one-to-one correspondence relation between theory and reality-in-itself is an idea constructed by anti-realists in order to diminish the possibilities of considering realism.} {\it Theories} relate on the one hand to {\it physis} (the One), and on the other to the multiplicity of phenomena (the Many). And it is this kernel aspect which marks the characteristic feature of scientific understanding itself. As a young Wolfgang Pauli would explain to his friend Werner Heisenberg during a conversation in 1921: 
\begin{quotation} 
\noindent {\small  ``knowledge cannot be gained by understanding an isolated phenomenon or a single group of phenomena, even if one discovers some order in them. It comes from the recognition that a wealth of experiential facts are interconnected and can therefore be reduced to a common principle. [...] `Understanding' probably means nothing more than having whatever ideas and concepts are needed to recognize that a great many different phenomena are part of coherent whole. Our mind becomes less puzzled once we have recognized that a special, apparently confused situation is merely a special case of something wider, that as a result it can be formulated much more simply. The reduction of a colorful variety of phenomena to a general and simple principle, or, as the Greeks would have put it, the reduction of the many to the one, is precisely what we mean by `understanding'. The ability to predict is often the consequence of understanding, of having the right concepts, but is not identical with `understanding'.'' \cite[p. 63]{Heis71}} 
\end{quotation}

But as we all know, very soon a strong opposition to these realist ideas also appeared in Athens, which had already become during the 5th. century B.C., the rich and prosperous capital of the Greek Empire. Sophists, as they would be called, would produce the first assault against the realist program arguing that there is no such thing as `a reality of things', and even if such thing would exist, we would not be able to grasp it. Sophists argued assuming a more down to earth position that we humans can only refer to our own perception. It makes no sense to talk about a reality independent of subjects because we individuals have only a {\it relative} access to things, an access limited by our own personal experience. As Protagoras would argue: ``Man is the measure of all things, of the things that are, that they are, of the things that are not, that they are not'' [DK 80B1]. From this skeptic standpoint, sophists criticized the physical idea of theoretical knowledge: realists ---namely, physicists and philosophers--- are too na\"ive, they do not recognize their own finitude and thus have ended up believing they can access the infinite, the real. The first battle of the war between realists and anti-realists had begun. 

As we know, it took the power of both Plato and Aristotle to overcome the sophist anti-realist riot. Through the creation of systems of thought rather than a reference to first elements ---in which Thales, Anaximines, Anaximandro and Empedocles had based their theories--- they were both able to provide an answer to the sophists and take the realist program a step further. In the specific case of Aristotle, metaphysics would imply an essential shift within the reference to {\it physis}, from `first elements' to a system of interrelated concepts and principles. Plato's and Aristotle's metaphysical systems were analyzed, discussed criticized and developed for two millennia keeping realist ideas in the center of the stage of Western thought. But with the rise of modernity in the 16th and 17th centuries things would begun to drastically change. The scientific revolution that took place in modernity might be regarded not only as a culmination of the Greek scientific path, but also as the beginning of its dissolution. On the one hand, the idea of the `One and the Many' was consistently articulated not only conceptually but also formally through the development of infinitesimal calculus. Physicists had finally constructed the first closed, unified, consistent and coherent formal-conceptual representation of physical reality, namely, classical mechanics. Through its development, physicists were capable of producing not only a conceptual qualitative understanding of phenomena but also the quantitative capacity to compute their accurate prediction. While the notion of {\it invariance} captured, in mathematical terms, a consistent unified representation of {\it the same} state of affairs independently of  {\it reference frames}. Objectivity captured an analogous content but now in purely conceptual terms, as a {\it moment of unity} defined in conceptual and categorical terms which allowed subjects to agree about the same object of experience. In this way, modern physics was able to solve the Ancient Greek problem of movement. Both invariance and objectivity would provide a rigorous set of formal-conceptual conditions ---to which we will return in section 5--- for producing a subject detached representation of a real {\it state of affairs} accounting for identity within difference, for sameness within change. 

However, on the other hand, even though modernity marks a period of essential accomplishments and advancements for the realist project, the 16th and 17th centuries might be also considered as marking the beginning of the anti-realist era. The human perspective was ---once again--- beginning to be considered as the true fundament of knowledge. Both rationalism, with the cartesian {\it cogito}, and empiricism, with their direct reference to experience, would begin to philosophize placing the subject at the center of their considerations. In doing so they were also introducing an essential separation within reality itself. As Heisenberg explains in {\it Physics and Philosophy}: 
\begin{quotation} 
\noindent {\small ``The great development of natural science since the sixteenth and seventeenth centuries was preceded and accompanied by a development of philosophical ideas which were closely connected with the fundamental concepts of science. It may therefore be instructive to comment on these ideas from the position that has finally been reached by modern science in our time. The first great philosopher of this new period of science was Ren\'e Descartes who lived in the first half of the seventeenth century. Those of his ideas that are most important for the development of scientific thinking are contained in his {\it Discourse on Method}. On the basis of doubt and logical reasoning he tries to find a completely new and as he thinks solid ground for a philosophical system. He does not accept revelation as such a basis nor does he want to accept uncritically what is perceived by the senses. So he starts with his method of doubt. He casts his doubt upon that which our senses tell us about the results of our reasoning and finally he arrives at his famous sentence: {\it `cogito ergo sum.'} I cannot doubt my existence since it follows from the fact that I am thinking. After establishing the existence of the I in this way he proceeds to prove the existence of God essentially on the lines of scholastic philosophy. Finally the existence of the world follows from the fact that God had given me a strong inclination to believe in the existence of the world, and it is simply impossible that God should have deceived me. This basis of the philosophy of Descartes is radically different from that of the ancient Greek philosophers. Here the starting point is not a fundamental principle or substance, but the attempt of a fundamental knowledge. And Descartes realizes that what we know about our mind is more certain than what we know about the outer world. But already his starting point with the `triangle' God-World-I simplifies in a dangerous way the basis for further reasoning. The division between matter and mind or between soul and body, which had started in Plato's philosophy, is now complete. God is separated both from the I and from the world. God in fact is raised so high above the world and men that He finally appears in the philosophy of Descartes only as a common point of reference that establishes the relation between the I and the world. While ancient Greek philosophy had tried to find order in the infinite variety of things and events by looking for some fundamental unifying principle, Descartes tries to establish the order through some fundamental division. But the three parts which result from the division lose some of their essence when any one part is considered as separated from the other two parts. If one uses the fundamental concepts of Descartes at all, it is essential that God is in the world and in the I and it is also essential that the I cannot be really separated from the world. Of course Descartes knew the undisputable necessity of the connection, but philosophy and natural science in the following period developed on the basis of the polarity between the {\it `res cogitans'} and the {\it `res extensa,'} and natural science concentrated its interest on the {\it `res extensa.'} The influence of the Cartesian division on human thought in the following centuries can hardly be overestimated, but it is just this division which we have to criticize later from the development of physics in our time.'' \cite[pp. 41-42]{Heis58}} 
\end{quotation}

The main accomplishment of the so called Enlightenment period ---which marks the starting point of the anti-realist approach to science \cite{deRonde20d}--- is the separation of {\it physis} in three different regions that would become complex in themselves and difficult to interrelate. It is this dissection that would allow in later times to destroy completely the meaning and content of the notion of reality. {\it Divide et impera}. This is the motto that might best characterize the anti-realist strategy against realism that begun in Modernity. Physics and philosophy were also torn apart and while the first was reassigned the specific role of discussing about the material world, the latter would be confined to the cogito and the subject. Ren\'e Descartes would ``cut'' the Greek notion of reality ---the One--- into three separated ``realities'': that of the I ({\it res cogitans}), that of the world ({\it res extensa}) and that of God. Very soon the new architectonic designed by the physicist and philosopher Immanuel Kant would take the reality of the Cartesian God ---which secured a correspondence relation between {\it res cogitans} and {\it res extensa}--- further away from the reach of science, into an un-knowable {\it noumenic} dimension. In Kant's co-relational metaphysics, reality was finally detached from scientific knowledge and replaced by the subject's capacity to account for objects of experience. The circular co-relation between subject and object had lost its foundation. Realism had been deathly wounded. Kant argued that the sum of all objects, the empirical world, is a complex of appearances whose existence and connection occur only in our {\it representations}. Reality, renamed as a beast, {\it das Ding an sich} (the Thing-in-Itself), had survived but only as a monstrous paradoxical creature hiding beyond empirical sensibility, impossible to be known. As Kant would write in the {\it Prolegomena to Any Future Metaphysics}: ``And we indeed, rightly considering objects of sense as mere appearances, confess thereby that they are based upon a {\it thing-in-itself}, though we know not this thing as it is in itself, but only know its appearances, viz., the way in which our senses are affected by this unknown something.'' Kant had introduced the un-knowable within his metaphysical system, limiting the scientific knowledge of physics to that of {\it objective reality} ---a reality restricted by his list of {\it categories} (grounded on Aristotelian metaphysics) and {\it forms of intuition} (Newtonian space and time) common to all human subjects. Thus, {\it noumenic} reality or reality-in-itself could not be considered anymore as the main goal of the scientific project. Friedrich Jacobi [1787: 223] famous remark would expose the problem in all its depth: ``Without the presupposition [of the `thing in itself,'] I was unable to enter into [Kant's] system, but with it I was unable to stay within it.'' Furthermore, Arthur Schopenahuer would make clear that the category of {\it causality} could not be applied within Kant's system to {\it noumenic reality} and consequently, the disconnection from the categorical representation of objective phenomena was complete. Kant had introduced an essential separation between theoretical representation and reality, changing the focus to subjects and objects. Reality had been cut into pieces and its essential unity finally captured and isolated. But it was still too soon for anti-realism to claim victory. Anti-realists would still have to wait two more centuries in order to rise as the supreme indisputable power of Western thought. It is in our postmodern age, during the 20th century, that realism would be finally defeated by anti-realism. And the main field of this final battle ---between realists and anti-realists--- would be no other than a new physical theory called quantum mechanics.

\section{Anti-Realist Realism, Myths and Quantum Mechanics}

According to Bas van Fraassen \cite[p. 2]{VF02}: ``Kant exposed the illusions of Reason, the way in which reason overreaches itself in traditional metaphysics, and the limits of what can be achieved within the limits of reason alone. [...] About a century later the widespread rebellions against the Idealist tradition expressed the complaint that Reason had returned to its cherished Illusions, if perhaps in different ways.'' By the end of the 19th century the Austrian physicist and philosopher Ernst Mach would produce a vigorous attack against the realist metaphysical presuppositions of classical mechanics. His development generated a new positive scheme for physics which, grounded on empirical observability alone, would attempt to finally erase metaphysics from physics. The attack was focused in the Newtonian notions of space and time ---which in the Kantian architectonic acted {\it a priori forms of intuition}--- and the notion of atom ---a metaphysical ancient creation by Leucippus and Democritus. The Austrian physicist and philosopher was breaking the walls of the modern spatiotemporal atomist cage in which physics had been confined. But for him, the destruction of this prison implied the necessary demolition of metaphysics itself.\footnote{It is interesting to note that concomitant with Mach's deconstruction of the metaphysics of classical physics, Friedrich Nietzsche would also produce a major attack to metaphysics within philosophy.} The deconstruction of Newtonian mechanics and Kant's metaphysical system would produce a major crisis in the foundations of science during the 20th century which Wolfgang Ernst Pauli ---the godson of Mach--- would recognize in his own terms: 
\begin{quotation} 
\noindent {\small ``In many respects the present appears as a time of insecurity of the fundamentals, of shaky foundations. Even the development of the exact sciences has not entirely escaped this mood of insecurity, as appears, for instance, in the phrases `crisis in the foundations' in mathematics, or `revolution in our picture of the universe' in physics. Indeed many concepts apparently derived directly from intuitive forms borrowed from sense-perceptions, formerly taken as matters of course or trivial or directly obvious, appear to the modern physicist to be of limited applicability. The modern physicist regards with scepticism philosophical systems which, while imagining that they have definitively recognised the {\it a priori} conditions of human understanding itself, have in fact succeeded only in setting up the {\it a priori} conditions of the systems of mathematics and the exact sciences of a particular epoch.'' \cite[p. 95]{Pauli94}} 
\end{quotation} 
Mach's subversive deconstruction had broken physics from its space-time atomist chains.\footnote{Mach's positivist scheme for science can be resumed in four main principles. The first is the naive empiricist idea that observation is a self evident {\it given} of ``common sense'' experience. Second, that physics should be understood as an economy of such observations. Third, that metaphysics, understood mainly as narratives about the unobservable, should be completely erased from scientific theories. And fourth, that physics does not talk about an ``external reality'' that would describe things beyond empirical observation.} And it was certainly this liberation which would become essential for the development of both Quantum Mechanics (QM) and relativity theory. While Albert Einstein applied Mach's positivist ideas in order to critically address the definition of {\it simultaneity} in classical mechanics, both Max Planck and Werner Heisenberg were able to advance new non-classical mathematical postulates and formalisms which could explicitly escape ---thanks to Mach's work--- the modern space-time representation of classical physics and in this way provide a quantitative operational account of a new field of (quantum) phenomena. However, even though positivist ideas were becoming popular in Europe ---specially among physicists--- Mach had lost a kernel battle against metaphysical atomism. Regardless of their endorsement to Mach's criticisms, physicists simply could not give up on the modern spatiotemporal representation of reality. As a consequence, even the new theory of quanta which had been developed through a radical departure from classical ideas and presuppositions was anyhow pictured as related to a microscopic realm constituted by elementary particles. The fact that Planck's {\it quantum postulate} precluded a continuous description did not seem to matter. The hope was that ---sooner than later--- this discreteness would be ---somehow--- explained in terms of a classical continuous representation. It is in this crossroad between quantum and classical that Niels Bohr ---maybe the most influential physicists of the 20the century--- would play an essential role establishing a new scheme for physics where Mach's positivist ideas would merge with a substantialist narrative detached from metaphysics. We will call this inconsistent but highly effective approach created by Bohr `anti-realist realism'. 

Bohr's spectacular appearance within the international physics community dates back to 1913 when he proposed a paradoxical quantum model for the Hydrogen atom. Even though Bohr had applied a strange mix of classical and quantum rules in order to reach predictive capacity, his proposal was framed within a straightforward classical narrative. The image he had created was simple and comforting to most physicists: electrons moved in quantized orbits around the nucleus just like planets orbited the sun. The microscopic realm appeared then as nothing but a reflection of our own planetary system. However, the price to pay for introducing this ``planetary narrative'' was plain formal and conceptual inconsistency ---extensively discussed within the specialized literature (see e.g., \cite{Vickers08} and references therein). From a conceptual perspective, it is clear that the existence of {\it discrete} quantum orbits within a {\it continuous} space simply did not make sense. How could continuous space be described in discrete terms? Why did electrons follow trajectories within confined orbits but were unable to reach the outer forbidden regions of space? How could they actually disappear from an orbit and reappear in another one without describing a trajectory? Where were electrons supposed to go during this apparently magical process? Furthermore, given that charged electrons were describing circular orbits, why didn't they irradiate, loose energy and collapse to the nucleus? From a formal perspective the inconsistency was even more explicit. Planck's discrete representation of energy implied, through the formula $E = \frac{m}{2} v^2$ (where $v = \frac{dx}{dt}$), that space and time could not be represented in continuous terms. If {\it energy} was ---according to Planck--- fundamentally {\it discrete} ($\Delta E$) then {\it velocity} had to be discrete as well, and consequently, also {\it space} and {\it time} ($v = \frac{\Delta x}{ \Delta t}$). Quantum discreteness was everywhere. And since the idea of a {\it discrete} space (or time) is essentially inconsistent, an {\it oximoron},\footnote{We might recall that the essential step for the development of the classical space-time representation of Newtonian mechanics was the creation of {\it infinitesimal calculus} which allowed for a rigorous mathematical definition of the {\it continuum}.} Planck's {\it quantum postulate} seemed to imply the birth of a new physics detached from the continuous space-time representation inherited from modern science. Unfortunately, this departure, in part due to Bohr's influence, would be never fully accepted neither by physicists nor philosophers. Against the radicalness of Planck's postulate, Bohr was able to reintroduce classical images and pictures in his model through a reinterpretation of the quantum of action in terms of an {\it irrepresentable} ``quantum jump'' taking place within the measurement interaction between quantum and classical systems. According to him \cite[vol. 1, p. 53]{Bohr87}: ``[the] essence [of quantum theory] may be expressed in the so-called quantum postulate, which attributes to any atomic process an essential discontinuity, or rather individuality, completely foreign to the classical theories and symbolized by Planck's quantum of action''. There is thus \cite[vol. 2, p. 61]{Bohr87} an ``essential ambiguity involved in a reference to physical attributes of objects when dealing with phenomena where no sharp distinction can be made between the behavior of the objects themselves and their interaction with the measuring instruments.'' It is through this gap within theoretical representation, that he himself had introduced, that he was able to picture the quantum realm in classical terms, gaining the sympathy of many conservative physicists who were not ready to give up on their ``commonsensical'' space-time atomist way of thinking. Bohr's model did not provide a theoretical account of the microscopic realm it supposedly described, but this was blamed on the theory itself which ---anyhow--- touched reality in the limits of representation. And in this respect, the classical images could be also regarded as ``just a way of talking''. But that is what physicists wanted to hear, and that is exactly what Bohr was giving them. Today, more than one century has passed by, but the strength and persistence of Bohr's account of QM continues to play a kernel role which allows physicists and philosophers to affirm the existence of quantum particles while at the same time deny the possibility of their consistent theoretical description.\footnote{Today, as Alisa Bokulich \cite{Bokulich11} recognizes: ``As we know well today [...] Bohr orbits are fictions ---according to modern quantum mechanics the electron in an atom does not follow a definite classical trajectory in a stationary state and is instead better described as a cloud of probability density around the nucleus.'' However, the notion of `particle' in contemporary physics, independently of its fragmentation \cite{Wolchover20}, continues to be regarded as essential \cite{deRondeFM21}.} In order to understand how all this became possible we need to go deeper into Bohr's complex scheme.  

%Bohr would generate through the introduction of {\it ad hoc} rules and principles a (anti-)system grounded on vagueness and inconsistency that would lead to a radical fragmentation in understanding.

In the first place, Bohr's {\it correspondence principle} \cite[p. 86]{Bohr63} would impose an ambiguous relation between classical physics and the quantum realm as ``the asymptotic approach of the description of the classical physical theories in the limit where the action involved is sufficiently large to permit the neglect of the individual quantum.'' Correspondence goes clearly in line with Bohr's understanding of QM as a rational generalization of classical mechanics (see \cite{Bokulich05}). Indeed, as he \cite[p. 87]{Bohr63} would later remark: ``the aim of [the correspondence principle] was to let a statistical account of the individual quantum processes appear as a rational generalization of the deterministic description of classical physics.'' Once again, even though there was no theoretical account of this limit between quantum and classical, the image was powerful enough to serve its purpose and restrict why certain transitions between stationary states occurred and others did not. However, regardless of Bohr's success with his model, there were still critical voices to be heard. Arnold Sommerfeld \cite{BokulichBokulich20}, one of the most prominent atomic physicists of the time, would argue: ``Bohr has discovered in his principle of correspondence a magic wand (which he himself calls a formal principle), which allows us immediately to make use of the results of the classical wave theory in the quantum theory.'' Some years later his criticism would grow stronger: ``The magic of the correspondence principle has proved itself generally through the selection rules of the quantum numbers, in the series and band spectra? Nonetheless I cannot view it as ultimately satisfying on account of its mixing of quantum-theoretical and classical viewpoints.'' As remarked by Alisa and Peter Bokulich \cite{BokulichBokulich20}: ``Sommerfeld's critical attitude toward the correspondence principle would prove influential on Wolfgang Pauli and Werner Heisenberg, both of whom were his doctoral students.'' Pauli would write to Bohr in a letter dated December 31st, 1924: 
\begin{quotation}
\noindent {\small ``I personally do not believe, however, that the correspondence principle will lead to a foundation of the rule? For weak men, who need the crutch of the idea of unambiguously defined electron orbits and mechanical models, the rule can be grounded as follows: If more than one electron have the same quantum numbers in strong fields, they would have the same orbits and would therefore collide.'' \cite{BokulichBokulich20}}
\end{quotation}
Also Heisenberg, who was at first clearly impressed by Bohr's correspondence program, would end up ---thanks to Pauli--- abandoning this ``magical'' guidance in 1925. Detaching himself completely from the classical attempt to describe the trajectory of electrons and applying instead Mach's observability principle as a standpoint of analysis, Heisenberg would finally reach a closed mathematical formalism capable to account for the intensive line-spectra observed in the lab in a consistent operational-invariant manner. For obvious reasons he would end up calling his new formulation Quantum Mechanics (QM). Unfortunately, Heisenberg's theory would be replaced ---just six months later--- by Schr\"odinger's wave mechanics which promised to restore a classical representation that was never delivered. The many problems encountered within Schr\"odinger's formulation would then open the doors ---four years later--- to the Dirac-von Neumann's axiomatic re-formulation which would finally impose Bohr's ideas in purely axiomatic terms.

\smallskip
\smallskip

\noindent {\sc Inconsistency of Quantum and Classical (Correspondence Principle):} {\it Quantum mechanics makes reference to a microscopic realm constituted by irrepresentable quantum particles. There exist a limit between this quantum microscopic realm and our classical macroscopic realm represented by classical physics.} 

\smallskip
\smallskip

At this point, some obvious questions might pop up to the attentive reader. First, how could Bohr introduce and make reference to a realm which could not be represented? If it couldn't be represented, then how did he know, for example, it referred to ``microscopic entities''? And how could there exist a limit or relation between a realm that couldn't be represented (i.e., the quantum) and another one (i.e, the classical) represented in terms of bodies existing within space and time? Essential to the effective censoring of these questions is Bohr's famous principle of {\it complementarity}. Shaped during his  debates with Albert Einstein during the 1920s, Bohr would become comfortable with the dualistic reference to `waves' and `particles' ---in the famous double-slit experiment--- in order to account for quantum objects in an inconsistent fashion. But it was in 1927 when, according to Heisenberg, Bohr came up with a more general understanding of his principle. Still two more years would have to pass by for complementarity to see the public light in Bohr's famous Commo lecture. As shown by Pekka Lahti \cite{Lahti80}, the principle of complementarity was not only vague and ambiguous but would metamorphosize itself through the years, relating to objects, properties and theories themselves.\footnote{For a detailed analysis  of the inconsistent nature of the complementarity principle see \cite{daCostaKrause06}.} Bohr's first step was to claim that in QM, we had lost the possibility to ``speak of the autonomous behavior of a physical object'' \cite[p. 87]{Bohr37}, and that as a consequence, ``[w]e must, in general, be prepared to accept the fact that a complete elucidation of one and the same object may require diverse points of view which defy a unique description.'' His conclusion \cite[vol.2, p. 40]{Bohr37} was that the ``evidence obtained under different experimental conditions cannot be comprehended within a single picture, but must be regarded as complementary.'' In this way, complementarity annihilated the conditions of possibility for any rational discourse about a common objective-invariant referent. As explained by Jean-Yves B\'eziau \cite{Beziau14}: ``[Bohr] argues that there are no direct contradiction: from a certain point of view `K is a particle', from another point of view `K is a wave', but these two contradictory properties appear in different circumstances, different experiments. Someone may ask: what is the absolute reality of K, is K a particle or is K a wave? One maybe has to give away the notion of {\it objective reality}.''  Bohr would rapidly extend his notion of complementarity in different directions from quantum objects (i.e., waves and particles) to properties (e.g., position and momenta) and then to the classical and quantum representations themselves (see \cite{Lahti80}). This extension was already explicit in Bohr's reinterpretation of Heisenberg famous inequality (see for a detailed analysis: \cite{HilgevoordUffink01}): 
\begin{quotation}
\noindent {\small ``Heisenberg's indeterminacy relations [...] specify the reciprocal latitude for the fixation, in quantum mechanics, of kinematical [position] and dynamical [momentum] variables required for the definition of the state of a system in classical mechanics. [...] in this context, we are of course not concerned with a restriction as to the accuracy of measurement, but with a limitation of the well-defined application of space-time concepts and dynamical conservation laws, entailed by the necessary distinction between [classical] measuring instruments and atomic [quantum] objects'' \cite[vol. 3, p. 5]{Bohr87}}
\end{quotation}
As he would also explain in his Warsaw lecture of 1938:
\begin{quotation}
\noindent {\small ``the statistical character of the uncertainty relations in no way originates from any failure of measurements to discriminate within certain latitudes between classically describable states of the object, but rather expresses an essential limitation of the applicability of classical ideas to the analysis of quantum phenomena. The significance of the uncertainty relations is just to secure the absence, in such an analysis, of any contradiction between different imaginable measurements.'' \cite[vol. 7, p. 311]{Bohr96}}
\end{quotation}
Later on, Bohr would famously argue in his reply to the EPR paper, that the value of complementary properties could not be considered as {\it prior} to the determination of the actual measurement set up:  
\begin{quotation}
\noindent {\small ``In the phenomena concerned we are not dealing with an incomplete description characterized by the arbitrary picking out of different elements of physical reality at the cost of sacrificing other such elements but with a rational discrimination between essentially different [complementary] experimental arrangements and procedures which are suited either for an unambiguous use of the space location or for an unambiguous use of the conservation theorem of momentum. Any remaining appearance of arbitrariness concerns merely our freedom of handling the measuring instruments, characteristic of the very idea of experiment. In fact the renunciation in each experimental arrangement of the one or the other of two aspects of the description of physical phenomena, ---the combination of which characterizes the methods of classical physics, and which therefore in this sense may be considered complementary to one another,--- depends essentially on the impossibility, in the field of quantum theory, of accurately controlling the reaction of the object on measuring instruments, i.e., the transfer of momentum in the case of position measurements, and the displacement in the case of momentum measurements. Just in this last respect any comparison between quantum mechanics and the ordinary statistical mechanics, ---however useful it may be for the formal presentation of theory,--- is essentially irrelevant. Indeed we have in each experimental arrangement suited for the study of proper quantum phenomena not merely to do with an ignorance of the value of certain physical quantities, but with the impossibilities of defining these quantities in an unambiguous way.'' \cite[p. 699]{Bohr35}}
\end{quotation}
Finally, another important extension of complementarity was provided in terms of the impossible relation between causal and classical space-time representations: 
\begin{quotation}
\noindent {\small ``On one hand, the definition of the state of a physical system, as ordinarily understood, claims the elimination of all external disturbances. But in that case, according to the quantum postulate, any observation will be impossible, and, above all, the concepts of space and time lose their immediate sense. On the other hand, if in order to make observation possible we permit certain interactions with suitable agencies of measurement, not belonging to the system, an unambiguous definition of the state of the system is naturally no longer possible, and there can be no question of causality in the ordinary sense of the word. The very nature of the quantum theory thus forces us to regard the space-time coordination and the claim of causality, the union of which characterizes the classical theories, as complementary but exclusive features of the description, symbolizing the idealization of observation and definition respectively. Indeed, in the description of atomic phenomena, the quantum postulate presents us with the task of developing a `complementarity' theory the consistency of which can be judged only by weighing the possibilities of definition and observation.'' \cite[p. vol. 1, pp. 54-55]{Bohr37}}
\end{quotation}

\smallskip
\smallskip

\noindent {\sc Inconsistency of Objects, Properties and Representations (Complementarity Principle):} {\it Quantum objects require contradictory classical representations provided through the notions of `wave' and `particle'. Complementary quantum properties (e.g., position and momentum) as well as measurement outcomes also require complementary experimental arrangements which are necessary as a prerequisite for their consideration. In QM space-time representations and causality are also complementary representations.} 

\smallskip
\smallskip

To sum up, the unity, consistency and coherency of theoretical representation, essential to the Greek scientific paradigm, developed also in modern times through the notions of {\it invariance} and {\it objectivity}, would become completely subverted within Bohr's matrix. The constitution of inconsistent dualities framed through {\it ad hoc} rules, principles and mythical pseudo-explanations would allow Bohr to create a new foundation for physics, shaky and unstable, constantly moving back and forth between waves and particles, position and momenta, causal mathematical representations and space-time events, between microscopic and macroscopic, between subjective observations and objective interactions, between classical ``common sense'' and a quantum irrepresentable realm, between reality and fiction... All this has been characterized by David Deutsch simply as `bad philosophy':
\begin{quotation}
\noindent {\small ``Let me define `bad philosophy' as philosophy that is not merely false, but actively prevents the growth of other knowledge. In this case [i.e., QM], instrumentalism was acting to prevent the explanations in Schr\"odinger's and Heisenberg's theories from being improved or elaborated or unified. The physicist Niels Bohr (another of the pioneers of quantum theory) then developed an `interpretation' of the theory which later became known as the `Copenhagen interpretation'. It said that quantum theory, including the rule of thumb, was a complete description of reality. Bohr excused the various contradictions and gaps by using a combination of instrumentalism and studied ambiguity. He denied the `possibility of speaking of phenomena as existing objectively'  ---but said that only the outcomes of observations should count as phenomena. He also said that, although observation has no access to `the real essence of phenomena', it does reveal relationships between them, and that, in addition, quantum theory blurs the distinction between observer and observed. As for what would happen if one observer performed a quantum-level observation on another, he avoided the issue. [...] For decades, various versions of all that were taught as fact ---vagueness, anthropocentrism, instrumentalism and all--- in university physics courses. Few physicists claimed to understand it. None did, and so students' questions were met with such nonsense as `If you think you've understood quantum mechanics then you don't.' Inconsistency was defended as `complementarity' or `duality'; parochialism was hailed as philosophical sophistication. Thus the theory claimed to stand outside the jurisdiction of normal (i.e. all) modes of criticism ---a hallmark of bad philosophy.'' \cite[p. 308-310]{Deutsch04}}
\end{quotation}

\section{A Smoky Dragon Flying in the Quantum Realm}

Maybe the most clear exposition of Bohr's approach to QM can be found in a paper by the prominent U.S. physicist John Archibald Wheeler ---not only one of Bohr's students but also one of his closest followers. In 1983 Wheeler co-authored with Warner Miller, a paper titled ``Delayed-Choice Experiments and Bohr's Elementary Quantum Phenomenon'' where they argued that the notion of {\it elementary quantum phenomenon} had to be regarded as the most important concept within the general scheme proposed by the Danish physicist.    
\begin{quotation}
\noindent {\small ``What one word does most to capture the central new lesson of the quantum? `Uncertainty', so it seemed at one time; then `indeterminism'; then `complementarity'; but Bohr's final word `phenomenon' ---or, more specifically, `elementary quantum phenomenon'--- comes still closer to hitting the point.  It is the fruit of his 28 year (1927-1955) dialog with Einstein, especially as that discussion came to a head in the idealized experiment of Einstein, Podolsky and Rosen. In today's words, no elementary quantum phenomenon is a phenomenon until it is a registered (`observed' or `indelibly recorded' phenomenon), `brought to a close' by an `irreversible act of amplification'.'' \cite[p. 72]{MillerWheeler83}}
\end{quotation}
But even though at first sight this notion might have seemed to many just a fancy way to talk about observations of `clicks' in detectors or `spots' in photographic plates, Wheeler had recognized that, in fact, there was a monstrous creature hiding beneath: 
\begin{quotation}
\noindent {\small ``The elementary quantum phenomenon is a great smoky dragon. The mouth of the dragon is sharp, where it bites the counter. The tail of the dragon is sharp, where the photon starts. But about what the dragon does or looks like in between we have no right to speak, either in this or in any delayed-choice experiment. We get a counter reading but we neither know nor have the right to say how it came. The elementary quantum phenomenon is the strangest thing in this strange world.'' \cite[p. 73]{MillerWheeler83}}
\end{quotation}
Of course, the smoky dragon ---contrary to Wheeler's account--- does not actually bite the counter, instead (as shown in figure 1) the `click' in the detector is direct consequence of its magical fire. Regardless of this obvious inaccuracy, Wheeler's dragon encapsulates perfectly well Bohr's approach to QM. An inconsistent scheme of mythical thought which has the main purpose of effectively justifying instrumental models through fictional concepts which even though have no theoretical nor experimental support are ---anyhow--- capable of upholding the most amazing illusions and narratives. A smoky dragon is a concept that cannot be (consistently) represented in theoretical terms, that has no experimental support but is anyhow regarded as describing an irrepresentable reality that becomes indistinguishable from fiction. 

\smallskip
\smallskip

\noindent {\sc Smoky Dragon (Inconsistent Mythical Concept):} A smoky dragon is an irrepresentable meaningless concept which provides a pseudo-picture of a physical situation or process and, consequently, the illusion of understanding. Such inconsistent concepts have no mathematical representation nor posses any operational testability procedure.

\smallskip
\smallskip

\begin{figure}
\centering
\includegraphics[scale=.1]{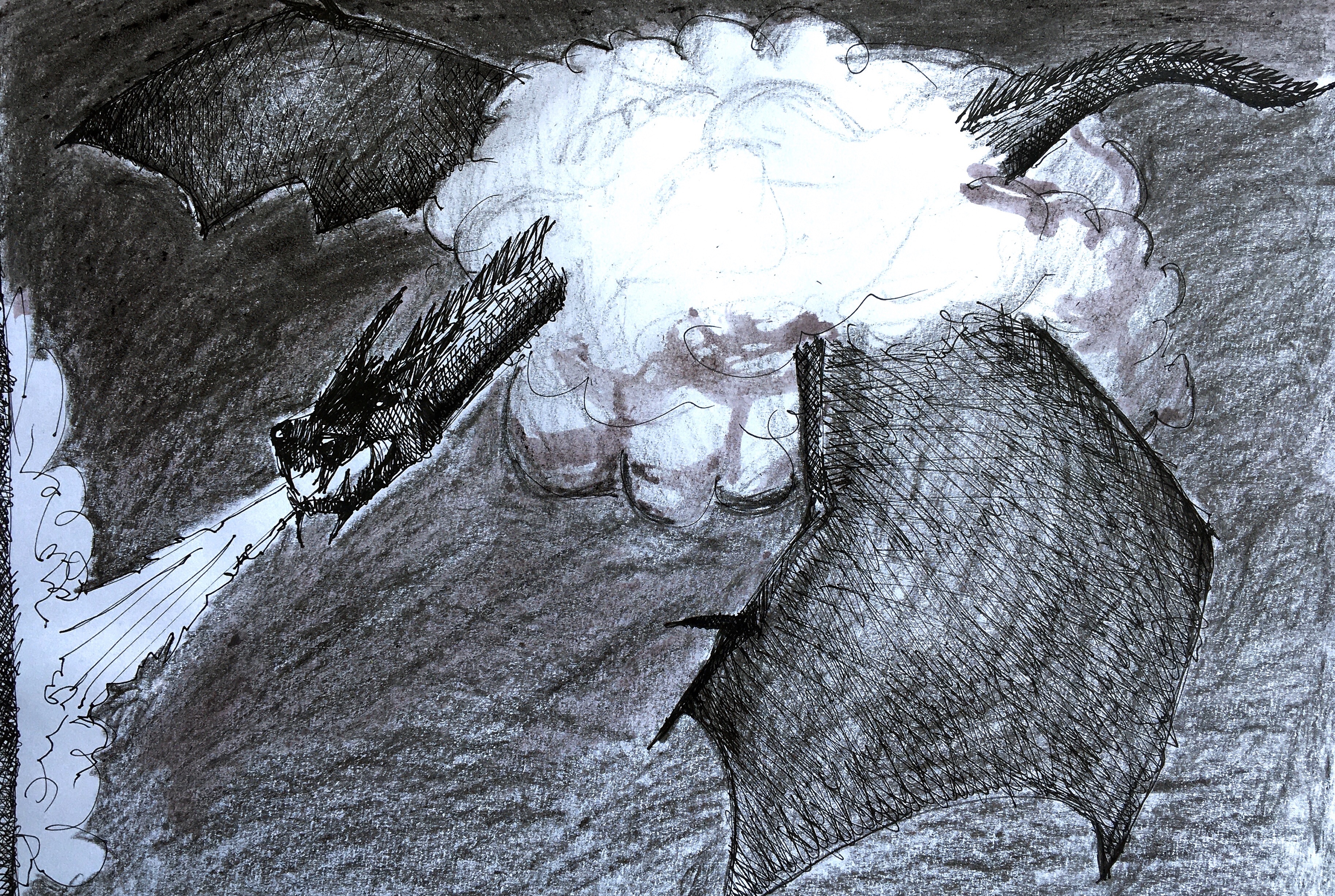}
\caption{ \small{Niels Bohr's mythical smoky dragon generating a `spot' in a photographic plate.}}
\end{figure}

\noindent Maybe the best example of a smoky dragon is Bohr's famous {\it quantum jump} of electrons between quantized orbits. For those acquainted with the theory, this notion generates a strange motion picture in our minds allowing us to imagine an impossible process that is not described by the mathematical formalism nor observed in the lab. Of course, hiding beneath these quantum jumps we find another smoky creature, namely, {\it quantum particles} themselves. Microscopic entities which must exist within space and time, since they are ``small'', but are anyhow impossible to represent. Quantum particles can be only witnessed through `clicks' in detectors and `spots' in photographic plates. The contradiction remains at plain sight, for if a quantum object is un-thinkable, irrepresentable, then ---obviously--- it cannot be ``small'' nor ``inhabit space and time''. Now, the question rises: how could such inconsistent fictional narratives endure within a ---supposedly--- rational field like physics? Essential to the survival of smoky dragons is Bohr's outstanding use of misdirection. Managing audience attention is the aim of all theater, and the foremost requirement of all magic acts. In theatrical magic, misdirection is a form of deception in which the performer draws audience attention to one thing to distract it from another. This is the key to understand the effectiveness of Bohr's prestiges. For example, in order to complete his atomic-planetary trick Bohr did not only rely on an atomist image that physicists where able to relate to something they already knew, he would effectively complement this ``commonsensical'' image with enigmatic ``quantum jumps'', a new process that would allow electrons to magically disappear from their orbit and immediately reappear in another one. The story was spectacular and physicists were immediately captured. How could this happen? What were these fantastic ``quantum jumps''? Were they actually real? How could particles disappear and reappear at will? Where were these particles going in the meantime? The complete lack of answers did not matter. The trick had been already performed. The Danish conjurer had succeeded in drawing the focus of attention away from the critical consideration of atoms, electrons and protons ---something that Mach had criticized just a few decades before--- to the fictional existence of ---unobservable and irrepresentable--- quantum jumps. Without theoretical and empirical support but with great confidence a young charismatic Bohr would whisper to his audience once and again: ``It is weird because it is quantum!'' 

It is interesting to notice that, regardless of its success and effectiveness, the complete lack of rational justification within Bohr's inconsistent scheme was already exposed in 1926, during a meeting that took place in Copenhagen where the Danish physicist had invited Schr\"odinger to discuss about the existence of ``quantum jumps'' within his new wave formulation of QM. Under the attentive gaze of Heisenberg, Bohr's young apprentice, the Austrian physicist would present many different arguments exposing not only the lack of theoretical and experimental support for the existence of quantum jumps but also the serious inconsistencies reached when introducing this phantasmagorical process within the theory. Schr\"odinger  \cite[p. 73]{Heis71} would then conclude that ``the whole idea of quantum jumps is sheer fantasy.'' At this point, with great mastery, without confronting the strong arguments of his opponent, Bohr would turn things completely upside-down in a single rhetorical move:
\begin{quotation}
\noindent {\small ``What you say is absolutely correct. But it does not prove that there are no quantum jumps. It only proves that we cannot imagine them, that the representational concepts with which we describe events in daily life and experiments in classical physics are inadequate when it comes to describing quantum jumps. Nor should we be surprised to find it so, seeing that the processes involved are not the objects of direct experience.'' \cite[p. 74]{Heis71}} 
\end{quotation}   
Reversing the burden of proof Bohr was asking Schr\"odinger either to grant him the existence of quantum jumps or prove their non-existence ---something known in jurisprudence as the {\it probatio diabolica}.\footnote{The legal requirement to achieve an impossible proof. Such Devil's Proof is the logical dilemma that while evidence will prove the existence of something, the lack of evidence fails to disprove it.} After his meeting, in a letter to Wilhelm Wien, Schr\"odinger would accept his defeat: 
\begin{quotation}
\noindent {\small ``Bohr's [...] approach to atomic problems [...] is really remarkable. He is completely convinced that any understanding in the usual sense of the word is impossible. Therefore the conversation is almost immediately driven into philosophical questions, and soon you no longer know whether you really take the position he is attacking, or whether you really must attack the position he is defending.'' \cite[p. 228]{Moore89}} 
\end{quotation}

\section{The Triumph of Anti-Realist Realism in 20th Century Physics}

In modern times, with the rise of co-relationalism, philosophy would engage itself in the clear cut dissection of the Greek notion of {\it physis} (section 1). As a culmination of this process, reality would become effectively separated in three distinct realms: subjective reality, objective reality and reality-in-itself. Kant would then limit physics to the circular interrelation between subjective and objective realities; distancing this co-relational form of knowledge from {\it reality-in-itself} which ---according to him--- would then remain unreachable, unknowable, unthinkable. Reality had been torn apart and detached from physics. Three centuries later, in post-modern times, Bohr was ready to generate a new system, more complex, stable and powerful than its predecessor. Through the introduction of fictions and irrepresentable referents Bohr would replace mathematical invariance, conceptual objectivity and operational testability ---to which we shall return in section 5--- by inconsistent principles and notions grounded on mythical narratives with no theoretical nor experimental support. The Bohrian matrix could be pictured as a highly effective M\"obius strip machine generating motion through the constant creation of dualistic poles applied within a never-ending line of reasoning. Going back and forth between contradictory statements and principles, Bohr was able to create a never-ending progression of rhetorical self-justification. Scrambling epistemology (i.e., gnoseology) with ontology he would argue that the fictional consideration of a quantum object imposed  a limit to representation itself \cite[v. 2, p. 62]{Bohr87}: ``In quantum mechanics, we are not dealing with an arbitrary renunciation of a more detailed analysis of atomic phenomena, but with a recognition that such an analysis is in principle excluded.'' It was the {\it quantum of action} which was to be blamed for this impossibility \cite[p. 79]{Bohr37}: ``not being any longer in a position to speak of the autonomous behavior of a physical object, due to the unavoidable interaction between the [quantum] object and the [classical] measuring instrument.'' Of course, according to Bohr, the discrete quantum theoretical representation of energy, $\Delta E = h.n$, proposed by Max Planck in 1900 was not only imposing a limit to the possibilities of representation, it was also describing something truly real, something going on in each and every interaction between a quantum particle and a classical apparatus. As he would explain in his reply to EPR: 
\begin{quotation}
\noindent {\small ``The impossibility of a closer analysis of the reactions between the [quantum] particle and the [classical] measuring instrument is indeed no peculiarity of the experimental procedure described, but is rather an essential property of any arrangement suited to the study of the phenomena of the type concerned, where we have to do with a feature of [quantum] individuality completely foreign to classical physics.'' \cite[p. 701]{Bohr35}} 
\end{quotation}   
Thus, (quantum) physics could not represent objective reality as described by quantum objects due to the (real) interaction between the (fictional) quantum object and the (real) classical measuring device. Twisting things once again, this epistemological limit had to be understood ---according to Bohr--- not as a technical limit of our instruments or human capabilities but rather as an ontological feature of reality-in-itself, namely, its own irrepresentability! This typical scrambling of gnoseological and ontological claims, of realist and fictional statements, is part of Bohr's incredibly effective pendular rhetorics. The Bohrian M\"obius strip of reasoning is an amazing device which forces us to remain in constant motion, always between two poles: between waves and particles, between the objective interaction of systems and subjective observations, between microscopic and macroscopic realms, between theory and measurement, between ontology and epistemology, between reality and fiction... 

Continuing the trend of thought imposed by Bohr during the first decades of the 20th century, the development of quantum physics after the war would lead to the final triumph of anti-realism over realism. As Karl Popper  \cite{Popper63} would famously depict the situation already in the late 1950s: 
\begin{quotation}
\noindent {\small ``Today the view of physical science founded by Osiander, Cardinal Bellarmino, and Bishop Berkeley, has won the battle without another shot being fired. Without any further debate over the philosophical issue, without producing any new argument, the {\it instrumentalist} view (as I shall call it) has become an accepted dogma. It may well now be called the `official view' of physical theory since it is accepted by most of our leading theorists of physics (although neither by Einstein nor by Schr\"odinger). And it has become part of the current teaching of physics.'' \cite[pp. 99-100]{Popper63}} 
\end{quotation}   
Popper would argue that this anti-realist outcome was a direct consequence not only of the successful technical applications obtained in the U.S., ``some of them with a big bang'', but also due to Bohr's complementarity: 

\begin{quotation}
\noindent {\small ``In 1927 Niels Bohr, one of the greatest thinkers in the field of atomic physics, introduced the so-called principle of complementarity into atomic physics, which amounted to a 'renunciation' of the attempt to interpret atomic theory as a description of anything. Bohr pointed out that we could avoid certain contradictions (which threatened to arise between the formalism and its various interpretations) only by reminding ourselves that the formalism as such was self-consistent, and that each single case of its application (or each kind of case) remained consistent with it. The contradictions only arose through the attempt to comprise within one interpretation the formalism together with more than one case, or kind of case, of its experimental application. But, as Bohr pointed out, any two of these conflicting applications were physically incapable of ever being combined in one experiment. Thus the result of every single experiment was consistent with the theory, and unambiguously laid down by it. This, he said, was all we could get. The claim to get more, and even the hope of ever getting more, we must renounce; physics remains consistent only if we do not try to interpret, or to understand, its theories beyond (a) mastering the formalism, and (b) relating them to each of their actually realizable cases of application separately. 

Thus the instrumentalist philosophy was used here {\it ad hoc} in order to provide an escape for the theory from certain contradictions by which it was threatened. It was used in a defensive mood to rescue the existing theory; and the principle of complementarity has (I believe for this reason) remained completely sterile within physics. In twenty-seven years it has produced nothing except some philosophical discussions, and some arguments for the confounding of critics (especially Einstein).'' \cite[pp. 100-101]{Popper63}} 
\end{quotation}   
John Clauser \cite[p. 70]{Clauser02} supports Popper's historical reading when he recalls the situation as a student in physics during the 1960s: ``given Bohr's strong leadership, the net legacy of their arguments is that the overwhelming majority of the physics community accepted Bohr's `Copenhagen' interpretation as gospel, and totally rejected Einstein's viewpoint.'' Another U.S. physicist, Henry Stapp \cite{Stapp72}, seems to agree when he recalls that ``the Copenhagen interpretation, was bitterly challenged at first but became during the '30's the orthodox interpretation of quantum theory, nominally accepted by almost all textbooks and practical workers in the field.'' As a direct consequence, questions about reality would become to be regarded as part of an external philosophical debate (see for a detailed historical analysis \cite{Freire15, Kaiser11}). As remarked by Ramirez \cite{Ramirez20} during the 1950s, 60s and 70s ``those physicists in the early moments of their careers who were interested in pursuing more foundational questions faced a strong opposition by the dominant academic physics culture of the time.'' And back to Clauser's analysis:\footnote{For analogous remarks by David Albert and Lee Smolin see \cite{deRonde20d}.} 
\begin{quotation}
\noindent {\small``Any physicist who openly criticized or even seriously questioned these foundations (or predictions) was immediately branded as a `quack'. Quacks naturally found it difficult to find decent jobs within the profession. [...] Religious zeal among physicists prompted an associated powerful proselytism of students. As part of the `common wisdom' taught in typical undergraduate and graduate physics curricula, students were told simply that Bohr was right and Einstein was wrong. [...] Any student who questioned the theory's foundations, or, God forbid, considered studying the associated problems as a legitimate pursuit in physics was sternly advised that he would ruin his career by doing so. I was given this advice as a student on many occasions by many famous physicists on my faculty at Columbia and Dick Holt's faculty at Harvard gave him similar advice.'' \cite[pp. 72-73]{Clauser02}} 
\end{quotation}   
Thus, as described in detail in \cite{Freire15, Kaiser11, Ramirez20}, QM in tune with the instrumentalist {\it Zeitgeist} would become to be taught during the 1960s and 1970s in all Universities around the globe simply as what Tim Maudlin has called an algorithmic ``recipe'' capable to predict measurement outcomes. Today, more than half a century later, the situation has not changed. As described by Maudlin \cite[pp. 2-3]{Maudlin19}: ``if a physics student happens to be unsatisfied with just learning these mathematical techniques for making predictions and asks instead what the theory claims about the physical world, she or he is likely to be met with a canonical response: Shut up and calculate!'' 

However, it is certainly not only naive instrumentalism which rules contemporary physics. It is the much more complex, effective and ambiguous anti-realist realism created by the Danish Nobel which guides scientific research. Still today, Bohr's pendular rhetorics about microscopic particles, classical apparatuses and measurement outcomes remains the discursive basis of ``Standard QM'' (SQM), namely, a general set of inconsistent postulates, ideas and principles framed under Dirac's and von Neumann's axiomatic formulation. Something also referred to ---by physicists--- as ``the Copenhagen interpretation'' or ---by philosophers--- as the ``minimal interpretation''. This general scheme is grounded on three widespread claims which share a profound inconsistent consensus between physicists. 
{\bf \begin{itemize}
\item[I.] QM describes microscopic particles.
\item[II.] QM cannot be understood. 
\item[III.] QM is only an algorithmic model which predicts measurement outcomes. 
\end{itemize}}
The first point of absolute agreement is the supposedly ``obvious fact'' that SQM makes reference to a microscopic realm constituted by elementary particles. As explained by the famous post-war U.S. physicist, Richard Feynman \cite[Chap. 37]{Feynman63}: ``Quantum Mechanics is the description of the behavior of matter and light in all its details and, in particular, of the happenings on an atomic scale.'' This does not imply, however, there is any consensus about what a quantum particle really is \cite{Wolchover20}. The second point of agreement between contemporary physicists can be resumed in another phrase made popular by Feynman \cite[p. 129]{Feynman67}:  ``nobody understands QM''. The impossibility to understand the theory is linked to another Bohrian premise, namely, that the un-intuitiveness of QM is linked to its departure from classical concepts and representation itself. 
\begin{quotation}
\noindent {\small ``It would be a misconception to believe that the difficulties of atomic theory may be evaded by eventually replacing the concept of classical physics by new conceptual forms. Indeed, [...] the recognition of the limitation of our forms of perception by no means implies that we can dispense with our customary ideas and their direct verbal expression when reducing our sense of impressions to order. No more is it likely that the fundamental concepts of the classical theories will ever become superfluous for the description of physical experience. The recognition of the indivisibility of the quantum of action, and the determination of its magnitude, not only depend on an analysis of measurements based on classical concepts, but it continues to be the application of these concepts alone that makes it possible to relate the symbolism of the quantum theory to the data of experience. At the same time, however, we must bear in mind that the possibility of an unambiguous use of these fundamental concepts solely depends upon the self-consistency of the classical theories from which they are derived and therefore, the limits imposed upon the application of these concepts are naturally determined by the extent to which we may, in our account of the phenomena, disregard the element which is foreign to classical theories and symbolized by the quantum of action.'' \cite[p. 16]{Bohr87}} 
\end{quotation}
Some decades later Feynman himself would take this same empiricist presupposition as a natural standpoint: 
\begin{quotation}
\noindent {\small ``In the beginning of the history of experimental observation, or any kind of observation on scientific things, it is intuition, which is really based on simple experience with everyday objects, that suggests reasonable explanations for things. But as we try to widen and make more consistent our description of what we see, as it gets wider and wider and we see a greater range of phenomena, the explanations become what we call laws instead of simple explanations. One odd characteristic is that they always seem to become more and more unreasonable and more and more intuitively far from obvious. [...] There is no reason why we should expect things to be otherwise, because the things of everyday experience involve large numbers of particles, or involve things moving very slowly, or involve other conditions that are special and represent in fact a limited experience with nature.'' \cite[p. 127]{Feynman67}} 
\end{quotation}
Finally, these two contradictory statements ---obviously, if we do not know what QM is talking about then we cannot say that it talks about atoms--- are supplemented by a fundamental instrumentalist claim according to which \cite{Bohr60}: ``Physics is to be regarded not so much as the study of something a priori given, but rather as the development of methods of ordering and surveying human experience.'' Bohr argued in this respect that: 
\begin{quotation}
\noindent {\small ``The entire formalism is to be considered as a tool for deriving predictions, of definite or statistical character, as regards information obtainable under experimental conditions described in classical terms and specified by means of parameters entering into the algebraic or differential equations of which the matrices or the wave-functions, respectively, are solutions. These symbols themselves, as is indicated already by the use of imaginary numbers, are not susceptible to pictorial interpretation; and even derived real functions like densities and currents are only to be regarded as expressing the probabilities for the occurrence of individual events observable under well-defined experimental conditions.'' \cite[p. 314]{Bohr48} }\end{quotation}
Once again, many decades later the physicists Chris Fuchs and Asher Peres \cite[p. 70]{FuchsPeres00} would continue to support Bohr's ideas by arguing that: ``[...] quantum theory does not describe physical reality. What it does is provide an algorithm for computing probabilities for the macroscopic events (`detector clicks') that are the consequences of experimental interventions.'' 

We can now begin to understand how through his smoky dragons Bohr was able to retain and ---at the same time--- subvert theoretical representation, scientific knowledge and reality itself. His replacement of (consistent) theoretical representations by (inconsistent) mythical stories would prove an effective strategy against an enemy shared with positivists, namely, metaphysical systems. Following {\it Michael Corleone's strategy:} ``Keep your friends close, and your enemies closer'' Bohr ---contrary to logical positivists \cite{Faye21}---, would avoid any direct confrontation with realists and metaphysicians who would be welcome to stay within his new territory, of course at the high cost of paying tribute to his more general anti-realist, anti-metaphysical (anti-)system. 

%PRECISION IS OF COURSE RELATIVE. IT IS COMPLETELY STUPID TO SAY THAT QM CAN MEASURE SOMETHING UP TO 14TH DECIMAL. WITH RESPECT TO WHAT MEASURE, METERS???? AND THAT IS NOT A CAPACITY OF THE THEORY BUT OF OUR MEASURING INSTRUMENTS YOU IDIOT!! THEORIES DO NOT HAVE ERRORS!!! 

%Smoky dragons are the building blocks of meaningless narratives, the magical elements of a fictional reference to an unreachable and unknowable reality which is capable to bite the counter.   

%A common way out of this conundrum is to argue that quantum particles are ``fundamental'' which is just a less aggressive way to say ``shut up and calculate!'' Indeed, as recognized by theoretical physicist at the Massachusetts Institute of Technology Xiao-Gang Wen \cite{Wolchover20}: ``We say [electrons, photons, quarks] are `fundamental'. But that's just a [way to say] to students, `Don't ask! I don't know the answer. It's fundamental; don't ask anymore'.'' 

\section{The ``Collapse Dragon'' and the ``Map of Madness''}

After the war, under the influence of Bohr's anti-realist realism, all systematic questioning about the relation between theory and reality would become banished from physics and placed under the jurisdiction of philosophy where ---not to our complete surprise--- we can also witness the profound influence of Bohr's anti-system. In fact, the creation of ``philosophy of QM'' as a discipline in its own right during the late 1970s and 1980s is linked to the most famous smoky quantum dragons of all times, namely, ``the collapse of the quantum wave function''. The story begins in the year 1930 when a young English engineer and mathematician named Paul Maurice Dirac, following the teachings of Bohr and logical positivists, would give birth to a dragon so powerful it could turn an abstract mathematical formula into an empirical observation ---or in more technical terms, a quantum superposition into a single measurement outcome. This would become to be known between physicists and philosophers as the ``collapse'' of the quantum wave function ---also called {\it projection postulate} or {\it measurement axiom}. The {\it ad hoc} addition of this fantastic process, an essential cornerstone of the standard contemporary understanding of the theory of quanta, was presented by Dirac within the first chapters of his book: {\it The Principles of Quantum Mechanics}. In a typical Bohrian fashion, making reference to the polarization measurement of photons, Dirac would describe the real effect of observations:
\begin{quotation}
\noindent {\small``When we make the photon meet a tourmaline crystal, we are subjecting it to an observation. We are observing whether it is polarized parallel or perpendicular to the optic axis. The effect of making this observation is to force the photon entirely into the state of parallel or entirely into the state of perpendicular polarization. It has to make a sudden jump from being partly in each of these two states to being entirely in one or the other of them. Which of the two states it will jump cannot be predicted, but is governed only by probability laws.'' \cite[p. 9]{Dirac74}} 
\end{quotation}

The introduction of this new ``quantum jump'' by Dirac is nothing more than a direct consequence of the premises and ideas behind Bohr's interpretation of QM. First, the metaphysical presupposition regarding the existence of a microscopic realm constituted by irrepresentable quantum particles (e.g., photons). Second, the complementary reference to classical ``common sense'' concepts (such as `particles', `waves' and consequently also `polarization') in order to describe this microscopic quantum realm. And third, the empiricist/instrumental reference to measurement outcomes ---`clicks' in detectors and `spots' in photographic plates--- as the only predictive content of the theory itself. Indeed, staying close to Bohr's atomist reference to a microscopic quantum realm, we find in the very first page of Dirac's book \cite[p. 1]{Dirac74} the following claim: ``it has been found possible to set up a new scheme, called quantum mechanics, which is more suitable for the description of phenomena on the atomic scale.'' In close agreement with Bohr's architectonic, Dirac would argue: ``We have, on the one hand, the phenomena of interference and diffraction, which can be explained only on the basis of a wave theory; on the other, phenomena such as photo-electric emission and scattering by free electrons, which show that light is composed of small particles.'' And applying Bohr's complementarity discourse he would conclude that: ``quantum mechanics is able to effect a reconciliation of the wave and corpuscular properties of light'' such that ``[a]ll kinds of particles are associated with waves in this way and conversely all wave motion is associated with particles. Thus, all particles can be made to exhibit interference effects and all wave motion has its energy in the form of quanta.'' Then, in a truly positivist manner, the English engineer and mathematician would also state that ``science is concerned only with observable things''. Finally, closing the Bohrian M\"obius strip, Dirac would emphasize the superfluous role of (metaphysical) conceptual representations: ``it might be remarked that the main object of physical science is not the provision of pictures, but the formulation of laws governing phenomena and the application of these laws to the discovery of phenomena. If a picture exists, so much the better; but whether a picture exists of not is a matter of only secondary importance.'' Thus, in analogous way to Bohr's change of focus, from the representation of atoms and electrons to irrepresentable orbital quantum jumps, Dirac would evade the conceptual questioning about quantum superpositions by creating a new misdirection, namely, the irrepresentable collapse process magically triggered by observation. While  orbital jumps had allowed Bohr to reinforce a metaphysical narrative about atoms, Dirac's observational jump allowed him to impose an actualist dogma according to which certainty would be restricted to a binary account of observations in which measurement outcomes would become the natural counterpart of quantum particles. In this way, the observational focus would radically shift from the intensive values discussed in matrix mechanics to the binary values grounded on the observation of single outcomes and microscopic particles. Quite ironically, Dirac's logical positivist account of physics ---which grounded the introduction of his actualist dogma--- was doing away with the positivist procedure which had allowed Heisenberg to come up with QM in the first place ---just five years before--- when considering what was actually observed in the lab, namely, intensive line spectra ---not measurement outcomes. Dirac's Bohrian moves were giving up consistency not only in the conceptual level of discourse but also in the mathematical level of representation. While the reference to binary values would preclude a global invariant account of states,\footnote{As discussed in detail in \cite{deRondeMassri17, deRondeMassri21}, even though value-invariance cannot be consistently established within the mathematical formalism of QM which presupposes a binary valuation, it is naturally restored globally when considering Heisenberg's original reference to intensive values.} the addition of the collapse would transform  the linear mathematical formalism of the theory into an inconsistent axiomatic system both linear and non-linear.\footnote{As remarked by Lewis \cite[p. 50]{Lewis16}: ``Measuring devices must obey the Schr\"odinger dynamics, since every physical system does, but they must also violate the Schr\"odinger dynamics if they are to enact the measurement postulate. Quantum mechanics, understood as including the measurement postulate, is not just incomplete; it is inconsistent.'' In  \cite[pp. 73-79]{Albert92} David Albert points to the same contradiction.} 

Dirac's Bohrian-positivist formulation was reinforced two years later by the influential mathematician John von Neumann and the publication of his {\it Mathematical Foundations of Quantum Mechanics} \cite{VN}. Standard QM had been born, but together with it, also the ``collapse dragon'' which very soon was ready to give birth to another monster baptized as ``the measurement problem of QM''. Indeed, the fact that the collapse of the superposition created a serious inconsistency within the mathematical formalism of the theory, that such a ``sudden jump'' had never ever been observed nor measured within the lab or that it was not even necessary from an operational perspective ---for Heisenberg's matrix mechanics already provided a consistent operational account of what was being observed \cite{deRonde20b}--- was not considered as relevant by those who followed the gospel of complementarity. Just like in Bohr's model, the goal to restore a classical atomist discourse about measurement outcomes had been accomplished even if that implied giving up formal-conceptual consistency and coherency. Even though the contradictions were all over the place, physicists did not seem to care. Only two realist rebels would refuse to obey the limits to understanding imposed by the Danish ruler. In complete solitude, Einstein and Schr\"odinger would denounce the serious problems implied by the collapse of the quantum wave function. Schr\"odinger would ironically argue:
\begin{quotation}
\noindent {\small``But jokes apart, I shall not waste the time by tritely ridiculing the attitude that the state-vector (or wave function) undergoes an abrupt change, when `I' choose to inspect a registering tape. (Another person does not inspect it, hence for him no change occurs.) The orthodox school wards off such insulting smiles by calling us to order: would we at last take notice of the fact that according to them the wave function does not indicate the state of the physical object but its relation to the subject; this relation depends on the knowledge the subject has acquired, which may differ for different subjects, and so must the wave function.'' \cite[p. 95]{Freire15}} \end{quotation}
Einstein would tell Everett \cite[p. 88]{Freire15} that he ``could not believe that  a mouse could bring about drastic changes in the universe simply by looking at it''.  And in a letter dated December 22, 1950 he would write to Schr\"odinger:  
\begin{quotation}
\noindent {\small ``You are the only contemporary physicist, besides Laue, who sees that one cannot get around the assumption of reality, if only one is honest. Most of them simply do not see what sort of risky game they are playing with reality ---reality as something independent of what is experimentally established. They somehow believe that the quantum theory provides a description of reality, and even a complete description; this interpretation is, however, refuted most elegantly by your system of radioactive atom + Geiger counter + amplifier + charge of gun powder + cat in a box, in which the $\Psi$-function of the system contains both the cat alive and blown to bits. Nobody really doubts that the presence or absence of the cat is something independent of the act of observation.'' \cite[p. 39]{ESPL}}
\end{quotation} 
As we all know, regardless of the many attempts of the two subversive realists to bring attention to the serious problems in which physics was being dragged, Bohr's reply to the EPR paper in 1935 would settle the debate once and for all. Something that ---sadly enough--- would delay the debate about quantum entanglement and its technological application for more than half a century.\footnote{As Jeffrey Bub \cite{Bub17} explains: ``Most physicists attributed the puzzling features of entangled quantum states to Einstein's inappropriate `detached observer' view of physical theory, and regarded Bohr's reply to the EPR argument (Bohr, 1935) as vindicating the Copenhagen interpretation. [...] it was not until the 1980s that physicists, computer scientists, and cryptographers began to regard the non-local correlations of entangled quantum states as a new kind of non-classical resource that could be exploited, rather than an embarrassment to be explained away.'' } After the war, as we described above, with the Bohrian-positivist alliance merging in the U.S. in the form of instrumentalism, the persecution of students willing to criticize the many inconsistencies of SQM would become a standardized practice inside the walls of physics' faculties.\footnote{This, of course, is a practice that continues in the present. As Sean Carroll \cite{Carroll20}, a popular U.S. physicist, explains: ``Many people are bothered when they are students and they first hear [about SQM]. And when they ask questions they are told to shut up. And if they keep asking they are asked to leave the field of physics.''} It would be only during the 1980s ---with Aspect's experiments proving the relevance of quantum entanglement and, consequently, also the metaphysical analysis produced by Einstein and Sch\"odinger during the 1930s--- that young physicists willing to abandon their scientific careers would be allowed to enter a completely new field of research called ``Philosophy of QM''. But very soon, also this new field would be constrained to solving the ``measurement problem'', namely, explain the existence of a collapse process apparently triggered by observation. Philosophers of QM would attempt to ``solve'' the problem by adding what in technical jargon philosophers called an ``interpretation'' of the theory ---a narrative that would explain what was going on (see \cite{deRonde20b}). Quite ironically, fixing the mess created by the anti-realist {\it ad hoc} addition of the ``observational jump'' was now regarded as an essentially  ``realist problem''. As Tim Maudlin \cite[p. 52]{Schlosshauer11} recently explained: ``The most pressing problem today is the same as ever it was: to clearly articulate the exact physical content of all proposed `interpretations' of the quantum formalism is commonly called the measurement problem, although, as Philip Pearle has rightly noted, it is rather a `reality problem'.'' But, was the measurement problem an oasis were realists could freely develop their critical ideas or a refugee camp where anti-realists could supervise and control the growing number of subversive rebels? Let's see.

As Herv\'e Zwirn \cite[p. 639]{Zwirn16} has recently described: ``Faced to what seems a real inconsistency inside the quantum formalism, physicists [and philosophers of physics] have proposed many solutions largely depending on their initial philosophical inclination.'' The addition of interpretations in order to do away with the inconsistencies found in QM has become the most common practice within philosophical debates about the theory. However, following Bohr's anti-realist realism this praxis has also become committed to the creation of vague and inconsistent mythical stories which, once again, has helped to change the focus of attention and evade the deeper debate about the underlying inconsistencies present in the Bohr-Dirac-von Neumann account of QM. And of course, with no objective systematic link between theory and interpretation, the multiplication of these narratives has found no limit (see for a detailed discussion \cite{Raoni20, RaoniJonas20} and references therein). David Mermin \cite[p. 8]{Mermin12} famously declared: ``[Q]uantum theory is the most useful and powerful theory physicists have ever devised. Yet today, nearly 90 years after its formulation, disagreement about the meaning of the theory is stronger than ever. New interpretations appear every day. None ever disappear.''\footnote{John Horgan  \cite[p. 88]{Horgan15} describes how he was confronted to this essential problem when attending in 1992 a symposium at Columbia University in which philosophers and physicists attempted to discuss the meaning of quantum mechanics: ``The symposium demonstrated that more than 60 years after quantum mechanics was invented, its meaning remained, to put it politely, elusive. In the lectures, one could hear echoes of Wheeler's it from bit approach, and Bohm's pilot-wave hypothesis, and the many-worlds model favored by Steven Weinberg and others. But for the most part each speaker seemed to have arrived at a private understanding of quantum mechanics, couched in idiosyncratic language; no one seemed to understand, let alone agree with, anyone else. [...] [W]hen I revealed my impression of confusion and dissonance to one of the attendees, he reassured me that my perception was accurate. ``It's a mess,'' he said of the conference (and, by implication, the whole business of interpreting quantum mechanics). The problem, he noted, arose because, for the most part, the different interpretations of quantum mechanics cannot be empirically distinguished from one another; philosophers and physicists favor one interpretation over another for aesthetic and philosophical ---that is, subjective--- reasons.''} Funny enough, the truth is that most working physicists ---which have been trained in an instrumentalist fashion--- do not even know that such an interpretational debate about QM even exists.\footnote{As Maximilian Schlosshauer \cite[p. 59]{Schlosshauer11} remarks: ``It is no secret that a shut-up-and-calculate mentality pervades classrooms everywhere. How many physics students will ever hear their professor mention that there's such a queer thing as different interpretations of the very theory they're learning about? I have no representative data to answer this question, but I suspect the percentage of such students would hardly exceed the single-digit range.''} And the few who know couldn't care less. The reason is simple, this ``realist debate'' is not regarded as part of the scientific enterprise. Science today is orthodoxly understood in empirical terms, something which van Fraassen \cite[pp. 202-203]{VF80} resumes as follows: ``an empiricist account of science is to depict it as involving a search for truth only about the empirical world, about what is actual and observable'', more specifically, ``science aims to give us theories which are empirically adequate: an acceptance of a theory involves as belief only that it is empirically adequate.'' This is why Roberto Torretti \cite[p. 367]{Torretti99} is aboslutely correct when he points out that interpretations of QM should be considered as ``meta-physical ventures [...] for they view the meaning and scope of QM from standpoints outside empirical science.'' Taking this point into consideration Arthur Fine \cite[p. 149]{Fine86} gives us what seems a very reasonable advise: ``Try to take science on its own terms, and try not to read things into science. If one adopts this attitude, then the global interpretations, the `isms' of scientific philosophies, appear as idle overlays to science: not necessary, not warranted and, in the end, probably not even intelligible.'' Of course, regardless of the sarcastic reference made by many anti-realists to ``interpretations'', their introduction has played a kernel role in the perpetuation of their own program by creating the perfect smoky shield for hiding the many dragons flying freely around the Bohr-Dirac-von Neumann standard formulation of QM which ---regardless of its many inconsistencies--- continues to be taught in all Universities around the globe. The creation of interpretations has allowed anti-realists ---who have, of course, infiltrated the realist debate--- to control the activities of the subversive rebels.\footnote{There are many anti-realists who have proposed their own ``interpretations'', e.g. van Fraassen has proposed a ``modal interpretation'', Fuchs has proposed the ``QBism interpretation'' and even Bohr's proposal is referred by physicists as ``the Copenhagen interpretation''.}  The problem is that, after a few decades of free creation of myths and fictions, the realist camp seems to have run completely out of control. The complete lack of experimental and theoretical constraints in order to restrict the creation of such interpretations ---tolerated in some cases and embraced with fanatism in others--- has turned philosophy of QM into a breeding field of smoky dragons. It is quite recently that the situation has become to be recognized as  ``untenable''. Ad\'an Cabello, a prominent physicist in the field of quantum information, has even characterized this interpretational field of debate as ``a map of madness'' \cite{Cabello17}. 

%THE NOTION OF EM WAVE IS NOT AN INTERPRETATION BUT AN ESSENTIAL CONTENT OF THE THEORY. 

According to our analysis, the situation we have described is the natural consequence of the Bohrian-positivist erasing of conceptual systems within physical theories and their dualistic understanding in terms of observations and mathematical models. It is the void left by conceptual systems which has encouraged the creation of mythical stories. The addition of interpretations has allowed to proclaim the tolerance towards inconsistency and vagueness, shifting the focus of attention away from the ungrounded algorithmic models which have essentially replaced the consistency, coherency and unity required by theories. Bohr's double thinking allows the conceptual level to enter the scene either as a ``common sense'' natural way to refer to observable entities such as tables and chairs (i.e., the manifest image) or as an un-intuitive scientific manner to refer through vague and inconsistent mythical narratives to reality-in-itself (i.e., the scientific image). In this way, the idea of metaphysics as a closed system of relational concepts is replaced by a pendular reference moving between ``common sense'' classical concepts and mythical creations capable of exposing the irrepresentability of reality-in-itself.  The foundation moves then back and forth between the ``common sense'' access to macroscopic objects which justifies the derivation of the microscopic realm and conversely, from a microscopic reality which grounds our experience of macroscopic objects themselves. A good example of this praxis is provided by Sean Carroll who argues in favor of the existence of parallel worlds in a multiverse: 
\begin{quotation}
\noindent {\small ``As you learn more and more about the world, as you do more science, as you uncover more and more facts, observations become more precise, you go into realms that you hadn't yet seen ---distant stars and galaxies and the subatomic world. Would you expect ---that as your learn all these new things--- your best description of the world would become more intuitive and everyday or more and more weird and surprising? It will become more and more weird and surprising because we are looking at things that we are not trained to experience.'' \cite{Carroll13}}
\end{quotation} 
Starting from the classical representation as an ``intuitive'' realm of experience and understanding (i.e., the manifest image), any narrative going beyond our ``common sense'' can be either tolerated or embraced as a possible description of an underlying yet irrepresentable state of affairs (i.e., the scientific image). Vagueness and inconsistency become then not just a limit to representation but a manifestation of quantum non-classicality itself. Since systematic consistency is not required and the link between experience and concepts restricted to ``the myth of the given'' there is no other possible outcome than the extreme fragmentation of interpretations. In this context, Bas van Fraassen ---one of the most prominent contemporary anti-realists--- explains that, regardless of the fact that any fictional story responding to the question `what is the world like according to the theory?' could ---in principle--- be {\it true}, the interpretation does not need to be actually true in order for the theory to be good \cite[p. 10]{VF80}. In fact ---according to anti-realists--- we will never know for real if any of these fictions describes truthfully reality-in-itself.\footnote{As Alan Musgrave \cite[pp. 1209-1210]{PS} remarks: ``As usually understood, the realism-antirealism issue centers precisely on the question of truth. Positivists deny the existence of `theoretical entities' of science, and think that any theory which asserts the existence of such entities is {\it false}. Instrumentalists think that scientific theories are tools or rules which are {\it neither true nor false}. Empistemological antirealists like van Fraassen or Laudan concede that theories have truth-values, even that some of them might be true, but insist that no theory should be {\it accepted as true}.'' It is only  what anti-realists have termed ``scientific realists'' who argue that interpretations are true.} Thus, it might seem far more reasonable and convenient to remain simply ``agnostic'' ---as van Fraassen himself has called his skeptic position. After all, empirical science is only meant to describe actual observations. Period. Realists are then easily portrayed by anti-realists either as ``naive'' or ``fanatic'' believers in myths they cannot justify nor relate to empirically adequate theories. Given we accept the orthodox characterization of empirical science, anti-realists are essentially correct.

\section{How to Capture and Defeat a Smoky Dragon}

It is essential to acknowledge that through its criticisms anti-realism has always played an essential role for the development of science. Without sophistry it would have been impossible for Plato and Aristotle to create metaphysics, without Hume it would have been impossible for Kant to understand the essential role of subjectivity within representation, and without Mach's criticism to Newtonian metaphysics it would have been impossible for Einstein and Heisenberg to develop relativity theory and QM. Undoubtedly, it is the balance between realism and anti-realism which has been always kernel for the critical advancement of science in the history of Western thought. Unfortunately for all of us, during the last century this balance has been broken and the distinction between realism and anti-realism annihilated. Anti-realists have been finally able to conquer both physics and philosophy producing the most outstanding re-foundation of science ---turning the primacy of theory over observation completely upside-down. At the same time they have subverted realism turning it into a bad joke of itself related either to God's eye or solipsism itself. So, in this context, is there any hope left for the reconstruction of a truly realist program of science that goes beyond mythical fictions and narratives? Or in other words, would it be possible for realists to capture and defeat smoky dragons? Let's see.

Smoky dragons are powerful illusions capable of creating the fantasy of an ungrounded reference with no theoretical nor experimental support. QM illustrates perfectly well the extreme dangers created by an extremist anti-realist account of science supplemented by supposedly realist fictions. The annihilation of conceptual critical thought produced by smoky dragons has created a desert in which realists wonder with no compass, preaching and fighting between each other for imaginary stories that no one really cares about. The power of these creatures comes from the darkness of un-scientific mythical thought exposing the return to a dark pre-scientific rationality. However, our optimistic claim is that realists already possess, buried in their own tradition, the tools and weapons to fight and defeat smoky dragons. Realists just need to dig them out and be ready for battle. They need to remember the basic ideas of a tradition of more than two millennia which made possible the scientific revolution that took place in modernity. Realists just need to remind themselves that theoretical representations are not interpretations, that physical concepts are not just words which allow us to tell a story and that, in physics, it is only the theory which decides what can be observed. According to realism, a physical theory allows for a unified formal-conceptual representation of a state of affairs which relates to experience coherently and consistently in a qualitative and quantitative fashion. In this respect, it is essential to note that, while physical representations are both dependent on (conceptual) subjective preconditions and (formal) reference frames, realism seeks to provide through objectivity and invariance a subject {\it detached} account of physical reality\footnote{This {\it subject detached} account of reality was also discussed by Pauli and Bohr in relation to the meaning of objectivity. See: \cite[p. 60]{Laurikainen88}.} where ``the distinction between `directly observable' and `not directly observable' has no ontological significance'' \cite[p. 175]{Dieks88a} for as remarked by Heisenberg: 
\begin{quotation}
\noindent {\small ``The history of physics is not only a sequence of experimental discoveries and observations, followed by their mathematical description; it is also a history of concepts. For an understanding of the phenomena the first condition is the introduction of adequate concepts. Only with the help of correct concepts can we really know what has been observed.''  \cite[p. 264]{Heis73}}
\end{quotation} 

\begin{dfn}
{\sc Realism:} The presupposition that {\it physis} (or reality) is knowable\footnote{It should be remarked , once again, that realism does not refer to ``things'' or ``stuff' ---whatever that might mean---, it refers to reality. Knowing reality does not imply the idea that representation should describe {\it reality-as-it-is}; i.e., the idea there should exist a {\it correspondence relation} between the representation of a theory and reality. This claim is actually made by many anti-realists (e.g., van Fraassen) in order to create a ``straw realist'' easy to attack and defeat.} through the creation of theories, namely, unified, consistent and coherent formal-conceptual invariant-objective representations which provide an account of a state of affairs and experience detached from both particular subjects and reference frames. 
\end{dfn} 
\begin{dfn}
{\sc Anti-Realism:} The claim that realists are wrong and that even if {\it physis} (or reality) would actually exist, due to our human limitations it would anyhow remain always un-reachable, un-knowable and irrepresentable. 
\end{dfn} 

What is essential to realize is that the mathematical formalism and the conceptual system that conform a theory are constrained by a specific set of {\it necessary conditions} within the realist program of physics. It is these same conditions, essential for the creation of adequate physical concepts, that could be used today as powerful weapons against smoky dragons. Let us begin by addressing two of these realist conditions in some detail. Operational-invariance and operational-objectivity provide a rigorous foundation for any theory to refer in a meaningful consistent manner to a real state of affairs that can be considered as detached from the perspectival perception of subjects or the choice of any mathematical frame of reference. {\it Operational-Invariance} points to the fact that a mathematical representation must be able to provide a consistent scheme for the operational testability considered with respect to different frames of reference (or perspectives). For any realist representation to be consistent, there must always exist an operational-invariant formalism which allows us to discuss what is really going on independently of the choice of the particular reference frame ---chosen in order to describe the state of affairs from a specific perspective. A rabbit running through a field might be described from a frame of reference attached to a speed-train or to the railway station, but even though both descriptions will be {\it different} (the values of physical quantities such as velocity and position will obviously differ in different reference frames) it is the consistency between these different descriptions which allow us to talk about {\it the same} rabbit.   
\begin{dfn}
{\sc Operational Invariance:} A physical concept must be able to provide a consistent unified account of its operational testability considered with respect to different frames of reference (or bases). 
\end{dfn}
This condition is fulfilled in classical mechanics via the Galielan transformations and in relativity theory via the Lorentz transformations. On the contrary, in QM the orthodox interpretation of probability in terms of binary measurement outcomes together with the non-invariant definition of quantum state has completely destroyed the operational invariance present within Heisenberg's original matrix formulation (see for a detailed analysis and discussion \cite{deRondeMassri17, deRondeMassri22}). This mathematical condition can be translated in conceptual terms as a operational-objective condition which imposes the need to discuss about {\it the same} object of experience independently of the particular viewpoints. Observers from different perspectives should be able to agree about what they observe. Thus, unlike in the case of the famous story by Jorge Luis Borges, {\it Funes the Memorius} \cite{Borges}, ``the dog at three fourteen (seen from the side) should have the same name as the dog at three fifteen (seen from the front)'' (see for a detailed analysis \cite{deRonde20a}).
\begin{dfn}
{\sc Operational Objectivity:} A physical concept must be able to bring into unity the multiple perspectives of a sameness as observed from different viewpoints. 
\end{dfn}
Classical mechanics and electromagnetism are good examples of how both the formal and the conceptual parts of a theory can be brought into unity in order to consistently and coherently imagine quantitatively and qualitatively the evolution of a state of affairs independently of empirical perspectives or reference frames. In these particular cases, the conceptual moments of unity that provide a consistent account of what is going on are {\it particles} and {\it electromagnetic waves}. This is of course precluded within the standard Bohr-Dirac-von Neumann formulation of QM where, instead of providing a unified account of phenomena, states are defined ---in close relation to George Berkeley's dictum, {\it esse est percipi}--- in terms of the singular observations of measurement outcomes, consequently, differing from themselves in each and every observation. Bohr's principle of complementarity comes then at the rescue in order to paste together inconsistent representations, properties and outcomes. Bohr's empiricist standpoint, in line with positivism \cite{Faye21}, contrasts radically with Einstein's theoretical realism according to which it is only the theory which decides what can be observed:
\begin{quotation}
\noindent {\small ``I dislike the basic positivistic attitude, which from my point of view is untenable, and which seems to me to come to the same thing as Berkeley's principle, {\it esse est percipi.} `Being' is always something which is mentally constructed by us, that is, something which we freely posit (in the logical sense). The justification of such constructs does not lie in their derivation from what is given by the senses. Such a type of derivation (in the sense of logical deducibility) is nowhere to be had, not even in the domain of pre-scientific thinking. The justification of the constructs, which represent `reality' for us, lies alone in their quality of making intelligible what is sensorily given.'' \cite[p. 669]{Einstein65}}
\end{quotation} 
This implies that the definition of a conceptual {\it moment of unity} (e.g., a wave or a particle) always precedes the understanding of experimental data. Furthermore, as famously remarked by Einstein when addressing the concept of {\it simultaneity}, a physical concept requires not only a mathematical and conceptual definition, it must also possess a clear operational content: 
\begin{quotation}
\noindent {\small ``The concept does not exist for the physicist until he has the possibility of discovering whether or not it is fulfilled in an actual case. We thus require a definition of simultaneity such that this definition supplies us with the method by means of which, in the present case, he can decide by experiment whether or not both the lightning strokes occurred simultaneously. As long as this requirement is not satisfied, I allow myself to be deceived as a physicist (and of course the same applies if I am not a physicist), when I imagine that I am able to attach a meaning to the statement of simultaneity. (I would ask the reader not to proceed farther until he is fully convinced on this point.)'' \cite[p. 26]{Einstein20}}
\end{quotation}
Thus, the development of an adequate physical concept ---unlike smoky dragons--- involves also the provision of a consistent link to its experimental testing. 
\begin{dfn}
{\sc Conceptual Operationality:} A physical concept must provide a consistent procedure which allows to operationally test if it is the case in a given situation. 
\end{dfn}
Most smoky dragons we have been discussing in the context of quantum theory, apart form their vagueness and lack of an explicit definition, fail to fulfill this very basic operational condition. The collapse dragon is a good example. As remarked by Dennis Dieks: 
\begin{quotation}
\noindent {\small ``Collapses constitute a process of evolution that conflicts with the evolution governed by the Schr\"{o}dinger equation. And this raises the question of exactly when during the measurement process such a collapse could take place or, in other words, of when the Schr\"{o}dinger equation is suspended. This question has become very urgent in the last couple of decades, during which sophisticated experiments have clearly demonstrated that in interaction processes on the sub-microscopic, microscopic and mesoscopic scales collapses are never encountered.'' \cite[p. 120]{Dieks10}}
\end{quotation}

However, as Dieks \cite{Dieks18} also acknowledges: ``The evidence against collapses has not yet affected the textbook tradition, which has not questioned the status of collapses as a mechanism of evolution alongside unitary Schr\"odinger dynamics.'' This detachment of QM with respect to experimental evidence has become part of mainstream theoretical physics where we find today some of the most important referents in the field arguing that experience should not be regarded as a necessary ingredient of physical theories and models, something particularly explicit in String Theory. In this respect, Nobel Laureate Gerard t' Hooft \cite{Hooft01} has recently defended the idea that: ``Working with long chains of arguments linking theories to experiment, we must be able to rely on logical precision when and where experimental checks cannot be provided.'' Following the same line of reasoning Steven Weinberg \cite{Weinberg03} has gone even further: ``I think 100 years from now this particular period will be remembered as a heroic age when theorists cut themselves temporarily free from their experimental underpinnings and tried and succeeded through pure theoretical reasoning to develop a unified theory of all the phenomena of nature.'' During the last years, Richard Dawid has continued to provide non-empirical arguments in order to justify the relevance of mathematical theories which are incapable of producing predictions that could be actually tested in a lab \cite{Dawid13}. 

The possibility to conceive and observe {\it the same} object of experience {\it changing} through time is an essential aspect of physics. Apart from providing its operational content physical concepts must also bring into unity the phenomena observed in different subsequent tests. Repeatability is an essential condition for physical research and analysis. Obviously, if every time we observe `something' it refers to `something {\it different}', it becomes then impossible to keep track of anything. Things which are observed only once are impossible to investigate from a scientific standpoint. This is a problem which is well known since Heraclitus' theory of becoming and was referred to by the Greeks as `the problem of movement'. In short, what is {\it the same} within {\it change}, what can be regarded as an {\it identity} within {\it difference}? Clearly, if there is no repeatability, the reference of different experiences is precluded right from the start and just like in {\it Funes the Memorious}, the necessary link between the observation of `a dog at three-fourteen' and `a dog at three fifteen' is nowhere to be found. The state of affairs is lost. Thus, any meaningful physical concept must be capable of providing the conditions of its testing in different instants of time in relation to {\it the same} underlying state of affairs.  
\begin{dfn}
{\sc Operational Repeatability:} A physical concept representing a moment of unity must be able to unify the multiple phenomena observed in different subsequent tests.  
\end{dfn}
An object is a conceptual machinery which must be capable of bringing into unity a multiplicity of different observations. An object which can only relate itself to a single measurement outcome is simply not an object, it is just an `event' which lacks the conditions required by any realist physical representation that can be considered as meaningful. Physics does not talk about single observations or events, it talks about  systems which evolve as described by a dynamic equation. Physics talks in terms of formal-conceptual invariant-objective representations of states of affairs which even though change and evolve remain {\it the same} in invariant-objective terms. In SQM, however, the notion of `quantum particle' ---according to the orthodox narrative--- particles are destroyed with every measurement that is actually performed (see for a detailed discussion \cite{deRonde20b}) and consequently do not fulfill {\it operational repeatability}. In fact, quantum particles are explicitly defined as existents which can be only observed once, they are irrepeatble `clicks' in detectors or `spots' in photographic plates. Every measurement creates always a new `click', a new `spot', a new `particle', different to the ones preceding it, different to the ones to come. Of course, this goes against the very basic goal of science which in the words of Pauli attempts to account, in conceptual terms, for the unity of different phenomena: ```Understanding' probably means nothing more than having whatever ideas and concepts are needed to recognize that a great many different phenomena are part of coherent whole.'' 

It is quite clear that the standard ``recipe'' of QM fails to fulfill any of the just mentioned basic conditions required for a realist understanding of the theory. It is these general set of conditions which should guide realists in their future production of a unified, consistent and coherent representation for the theory of quanta. In this respect, the unquestionable fact that QM is ``weird'' should not be understood as necessarily imposing a limit to physical representation, but rather as exposing the limits of the modern representation of physics. The Bohrian claim that experience can be only described in terms of classical concepts can be then understood as a dogma which goes clearly agains any development of the theory of quanta. If we accept, as any realist should, that experience is derived from the theory ---and is not the ``self evident'' unproblematic given which grounds science---, then we will be necessary confronted not to the inescapable irrepresentability of the theory of quanta but with the need to develop a new adequate conceptual scheme that fullflls the conditions we have discussed.

\section{Conclusion}

In this article we have argued that interpretations and narratives attempting to justify an anti-realist fictional collapse of the quantum wave function ---introduced in order to make reference to single measurement outcomes without any experimental or theoretical support--- are clearly non-starters for a realist account of QM. Fictions have nothing to do with realism. Instead, the realist program should focus itself in the attempt to produce an invariant-objective representation of a real state of affairs. Something that can only be done by following the general theoretical conditions we have discussed in the previous section. From a realist standpoint, we must recognize that smoky dragons are nothing but contemporary myths which embrace contextuality instead of objectivity, preferred bases instead of operational-invariance, measurement-collapses and outcomes instead of operational repeatability. The lesson coming from this analysis is quite straightforward. It is only by staying close to the basic ideas of realism, refusing to enter the anti-realist labyrinth of interpretations and narratives, that we can hope to capture and defeat these fantastic creatures. The fight against them has just begun...

\section*{Acknowledgements} 

I want to thank Raimundo Fern\'andez-Mouj\'an, Axel Axel Eljatib, Matias Graffigna and Juan Vila for discussions on subjects related to this paper. This work was partially supported by the following grants: Project PIO CONICET-UNAJ (15520150100008CO) ``Quantum Superpositions in Quantum Information Processing'', UNAJ INVESTIGA 80020170100058UJ.

%----------------------------------------


\begin{thebibliography}{9999}
\bibitem{deRonde20b} de Ronde, C., 2020, ``Measuring Quantum Superpositions (Or, ``It is only the theory which decides what can be observed.'')'', preprint. (quant-ph:2007.01146).
\end{thebibliography}
\end{document}